\documentclass[11pt]{article}
\usepackage{fullpage}
\usepackage{url}
\usepackage{xspace} 

\newcommand{\mytitle}[1]{\noindent\makebox[\textwidth]{\hrulefill}\vspace{1ex}
  \begin{center}{\large \bf #1}\end{center}}
\newcommand{\myauthor}[1]{\begin{center}#1\end{center}}

\newcommand{\mysec}[1]{\vspace{-1ex}\subsubsection*{#1}\vspace{-1ex}}

\newcommand{\myabs}{{\normalfont \bf (Abstract)}\hspace{-1ex}~}
\renewenvironment{abstract}{}{}

\usepackage{enumitem}
\setlist[itemize]{noitemsep, topsep=.75ex}

\usepackage{lipsum}
\let\OLDthebibliography\thebibliography
\renewcommand\thebibliography[1]{\vspace{-1ex}
  \OLDthebibliography{#1}
  \setlength{\parskip}{0pt}
  \setlength{\itemsep}{0pt plus 0.3ex}
  \small
}

\begin{document}

\title{AppLP: A Dialogue on Applications of Logic Programming}

\author{{David S. Warren ~~~~~~ Yanhong A. Liu}\\
Stony Brook University}
\date{}
\maketitle






This document describes the contributions of the 2016 Applications of
Logic Programming Workshop (AppLP), which was held on October 17 and
associated with the International Conference on Logic Programming
(ICLP) in Flushing, New York City.

\mysec{Focus and scope}

The focus of the workshop was applications of logic programming, i.e.,
application problems, in whole or in part, that are solved by using
logic programming languages and systems. A particular theme of
interest was to explore the ease of development and maintenance,
clarity, performance, and tradeoffs among these features, brought
about by programming using a logic paradigm.  The goal was to help
provide directions for future research advances and application
development.

Real-world problems increasingly involve complex data and logic,
making the use of logic programming more and more beneficial for such
complex applications.  Despite the diverse areas of application, their
common underlying requirements are centered around ease of development
and maintenance, clarity, performance, integration with other tools,
and tradeoffs among these properties.  Better understanding of these
important principles will help advance logic programming research and
lead to benefits for logic programming applications.

The workshop was organized around four main areas of application:
Enterprise Software, Control Systems, Intelligent Agents, and Deep
Analysis.  These general areas included topics such as business
intelligence, ontology management, text processing, program analysis,
model checking, access control, network programming, resource
allocation, system optimization, decision making, and policy
administration. The issues proposed for discussion included language
features, implementation efficiency, tool support and integration,
evaluation methods, as well as teaching and training.

\mysec{Program and organization}

The workshop program included four sessions, one for each area, and each
session included an invited talk followed by contributed short
talks.  Participants were invited to submit a position paper, of one or
two pages, explaining their application problems,
solutions, rationales, and challenges.  Besides an excellent submission
for an invited talk, six others were chosen by the program committee to
present their work in short talks.  They were encouraged to make their
talks accessible to non-specialists.

Others were invited to participate on panels.
The workshop featured
discussions in which the moderator directed a short question to
a panelist, who presented their thoughts, after which the other
panelists were given a brief time to respond and contribute their
thoughts.  At the end of the workshop, the floor was opened to the
audience for a final open discussion.

The workshop brought together researchers and software engineers who are
building applications using logic programming, and presented novel
and challenging work and participated in lively discussions. 
Approximately 35 people participated in the morning session and about 55 in the afternoon.
This document includes summaries of all presentations and discussions, 
followed by abstracts and position papers, following their order in the program:



\begin{itemize}
\item The morning consisted of two sessions and a panel discussion. 
Session 1 was titled ``Enterprise Software and More'', with an invited talk by
Molham Aref, and contributed talks by Jeffrey Rosenwald and by Paul Fodor for Iliano Cervesato and Edmund Lam.  Session 2 was ``Control Systems and More''
with an invited talk by Marcello Balduccini and a contributed talk by Allesandra Russo.
The panel discussion was on ``Concurrent and Distributed Systems, Integration'',
moderated by Warren, with panelists Manuel Hermenegildo, Boon Thau Loo, Theresa Swift and Jan Wielemaker.

\item The afternoon included two sessions, another panel discussion, and an open 
discussion.  Session 1 was titled ``Intelligent Agents'', with an invited talk by Francesco Ricca and a contributed talk by Gopal Gupta.  
The panel was on ``Knowledge and Constraint Systems, Integration'', moderated by Warren, with panelists Marc Denecker, Torsten Schaub, and Mirek Truszczynski.
(Michael Gelfond was to be a panelist but unable to participate at the last minute.)  
Session 2 was on ``Deep Analysis'', with an invited talk by C.R. Ramakrishnan, and contributed talks by Nikolaj Bj{\o}rner and by Paul Tarau.
The open discussion was on ``Directions for Research and Applications: Big Data Analysis in Depth and Scale''.
\end{itemize}

As chairs of AppLP, we were aided by a strong program committee who reviewed the submitted papers and provided helpful input on the program form.  Each paper was reviewed by at least three committee members. The committee included:
  
\begin{center}
\begin{tabular}{l l}
Manuel Hermenegildo & IMDEA Software Institute, Spain \\
Bob Kowalski        &       Imperial College London, UK \\
Nicola Leone         &       University of Calabria, Italy \\
Michael Leuschel     &    Heinrich-Heine University in Dusseldorf, Germany\\ 
Vladimir Lifschitz   &       University of Texas at Austin, USA \\
Enrico Pontelli      &        New Mexico State University, USA \\ 
Theresa Swift        &       NOVA LINCS, Universidade Nova de Lisboa, Portugal \\ 
Jan Wielemaker       &    Vrije University Amsterdam, Netherlands\\
\end{tabular}
\end{center}

Local organization was aided by Bo Lin, Christopher Kane, and Saksham Chand of Stony Brook University. 
We would also like to thank Michael Kifer and Neng-Fa Zhou for their help as general chairs of ICLP.
AppLP 2016 was greatly improved by the generous sponsorship of
LogicBlox, Inc.\ for enabling additional discussions at organized lunch and dinner.

\pagebreak

\newcommand{\mypar}[1]{\vspace{-1ex}\subsubsection*{#1}\vspace{-1ex}}

\setlist[description]{labelindent=2ex,itemsep=-.5ex}


\mytitle{Summaries of Presentations and Discussions}
\myauthor{Prepared by Christopher Kane, Stony Brook University}

\maketitle

\mypar{LogicBlox: Solver-Aided Declarative Programming --- Invited talk by Molham Aref}


This talk introduced the ``smart database'' system LogicBlox (developed by the company
of the same name). It is organized as follows: (1) background information on enterprise
systems and motivation for new systems like LogicBlox, (2) important features of their
language LogiQL, (3) a brief explanation of their methods for efficiently computing
queries, and (4) results comparing the performance of LogicBlox to other similar
systems on standard benchmarks.



LogicBlox's customers' businesses are complex systems. Simplifying models
of important domains 
in their businesses are required to manage this 
complexity and produce useful analysis. Current solutions, 
spreadsheets and enterprise systems, are too limited and coarse-grained to 
handle the complexity effectively. In addition, current enterprise systems are
formed of many components that must be carefully coordinated, making them hard 
to use and maintain. LogicBlox strives to offer a unified system to avoid
these issues.

The primary users of LogicBlox are domain experts who 
are best-suited to building models of their businesses, not application 
developers. The LogicBlox language, LogiQL, is an expressive, declarative 
language based on Datalog meant to be accessible to domain experts. The talk gives 
several examples illustrating how derivation rules can be built from 
relations using conjunction, disjunction, negation, as well as arithmetic 
operations, aggregation, recursion, and existential quantification. 
Also explained are integrity constraints, derivation rules
with false in the conclusion, that are used to constrain the possible state.

The LogicBlox approach to efficient computation is ``brains before brawn'', by using
the best possible algorithms and data structures, rather than brute
force and better hardware.
Two examples demonstrate this approach. The first is the use of incremental computation
to maintain materialized views and efficiently update the results of repeated
queries as the database changes. The second is the use of the Leapfrog
Triejoin, a worst-case optimal algorithm for computing multi-way joins.
Simultaneous multi-way joins provide asymptotically better results for complex
queries than sequences of pairwise joins. The algorithm has been further
improved by the addition of worst-case optimal query planning and the results
can be incrementally maintained.

LogicBlox is superior to similar systems
at solving graph queries as shown by results for the clique benchmark. For a graph
with one hundred thousand edges LogiBlox performed 33 times faster than
Redshift and 57 times faster than HANA. For a graph with ten million edges,
LogicBlox performed 227 times faster than Redshift, while HANA failed to
achieve a result.

\mypar{Logic Programming in the Materials Handling and Logistics Industries ---
Presentation by Jeffrey Rosenwald}

This talks begins by comparing software engineering in two very different
industries, materials handling--moving items around large warehouses--and 
telecommunications. The requirements for materials handling
software are the same as the requirements for telecommunications. Each domain
requires large, distributed software systems capable of handling large numbers 
of concurrent activities featuring complex functionalities. These systems must be
in continuous operation for many years with maintenance conducted while the 
system is in operation. The systems must meet strict quality and reliability 
guidelines, and display high-tolerance for both software and hardware failures.
However, 
telecommunications systems have historically achieved greater
success at meeting these requirements than materials handling systems.

This talk proposes that developers of materials handling systems take
some lessons from the use of Erlang by telecommunications developers:
making a process the locus of failure can prevent failure from spreading,
processes communicate only by message passing, and processes can come and go.
However, 
SWI-Prolog has several advantages over Erlang for
writing materials handling software. The talk describes an agent-oriented system
written in SWI-Prolog by building on a ``holarchy of holons''. A holon is a small,
stand-alone program that has a specific sphere of influence within the system.
Among other virtues, the system at twenty thousand lines is an order of
magnitude smaller than alternative systems written in Java.

Unfortunately, there is significant resistance to adopting a system written in
Prolog. This is due to a suspicion of open source
technologies in general, and, more specifically, an ignorance of Prolog in
U.S. industry. Old businesses like materials handling are not willing to take a
risk on a system built on such an unfamiliar basis.  As a result, this system has 
been used for fast prototyping of continuous extensions and new installations 
but not as an adopted system in production.

\mypar{Concurrent Logic Programming: Met and Unmet Promises ---
Presentation by Paul Fodor based on slides by Iliano Cervesato and
Edmund Lam}

This talk begins by describing the promise of logic programming. The  
declarative nature of logic programming promotes human-friendly descriptions of problems
making them easier to understand and reason about. Some of this promise has been 
realized, but logic programming still struggles with very large programs and 
expected simplification of reasoning has not appeared consistently. The
largest problem for logic programming is that it remains a
fringe paradigm.

It is argued in this talk that this last problem can be addressed by applying logic programming
to a high-profile problem: concurrent and distributed application development. 
This is a good choice because these applications are everywhere, they are
difficult to get right, and no other programming paradigm contains a widely
adopted solution. Logic programming is suited to specifying
communication and synchronization in distributed applications, and reasoning
about their correctness.

The language, Comingle (developed by Iliano Cervesato and
Edmund Soon Lee Lam) supports the writing of mobile, distributed applications for
Android and i386 in a system-centric way (rather than the error-prone
node-centric fashion common to other programming paradigms). The language
implements a portion of first-order logic as horn clauses, evaluated with a
forward-chaining semantics. Techniques for reasoning about Comingle
applications are still being developed using ``session types'' and
``coinductive methods'' to determine that these applications behave the way they
should. Initial results for application programming and reasoning are
promising.

\mypar{What Tweety-the-Penguin and Faulty Suitcases Tell Us about Productivity,
Cybersecurity, and Data Sciences --- Invited talk by Marcello Balduccini}

This talk surveyed applications that use logic programming for
knowledge representation, where knowledge representation has three components:
(1) commonsense and non-monotonic reasoning (NMR), (2) non-monotonic logics and
constraint satisfaction programming (CSP), and (3) reasoning about actions and change (RAC). NMR
allows new knowledge to invalidate previous conclusions. There are several
formalisms for capturing NMR, including Prolog and ASP. Common sense and NMR can
be used in ASP to capture RAC.

The first example application uses ASP to build an automated system, 
USA-Advisor, to provide decision-support for the Space Shuttle's Reaction Control
System (RCS). RCS is a complicated set of tanks, valves, jets, switches, circuits, 
plumbing, and computer commands that are used to maneuver the shuttle while it 
is in orbit. Given a maneuvering goal $G$, USA-Advisor
determines whether a given set of RCS actions will achieve $G$, and if so,
generates a plan to achieve $G$, in the presence of arbitrary mulitple faults
in RCS. Over many test instances the ASP-powered USA-Advisor was able
to efficiently find a solution: 11 seconds on average, 1-2 minutes for hard
cases.

The second example application is industrial-scale print shop 
scheduling. There are constraints on the scheduling, including resource 
consumption, device availability, job phases, new jobs, device failures, and 
heuristics.  ASP does not scale well enough to handle a problem this complex. 
Prolog is efficient enough, but the implementation of the scheduler is not 
nearly as simple and clear as the abstract logic of the scheduler.  The solution 
was a new language that combines ASP with CSP, called EZCSP. This provided 
the efficiency required for industrial problems, like print shop scheduling, 
but retains the clarity of an ASP implementation of the scheduler.

The third application is automated malware mitigation---the
process of removing malware from an infected system effectively and safely. This
requires substantial reasoning because of the complex interdependencies between
components of an infected system.  So a precise, formal definition of the 
mitigation task was required. This was done by modeling the 
infected system as an RAC problem. Doing so
reduces the mitigation problem to a planning problem in which one looks for a 
sequence of actions that lead to a safe (i.e., malware-free) state. Testing on a
1000 simulated systems, each infected with 1 to 5 instances of malware and offering 1-40 
essential services, yielded a success rate close to 90\%, with the solution found 
in less than 2 seconds on average.

The fourth application is action-based information 
retrieval. Search engines excel at retrieving information about discrete facts,
but do poorly at retrieving relevant information in response to complex queries
about events. The response must contain information about the causes and 
outcomes of events. This can be done by representing this information as an 
instance of RAC. Reasoning about actions can be used to retrieve information 
about event outcomes. A representation 
framework has been built, natural language processing of source documents was
used to build connections between documents that describe events, their causes,
and their outcomes, and action-based retrieval has been demonstrated in case 
studies; but much of this problem remains open.

The talk concludes that many practical applications can, and are, being 
developed using the combination of the three components illustrated for knowledge
representation.

\mypar{Distributed Systems Management: Logic Programming Solution and
Challenges --- Presentation by Alessandra Russo}

This talk presented an evaluation of efforts to apply logic programming to
distributed systems management, focusing on two applications: policies for access control and distributed
computing.

Access control policy languages must be expressive and provide a
good framework for analysis. Logic programming languages are expressive enough
to capture a range of existing policy languages and can provide
clear formalizations of both policies and system states. Abductive constraint
logic programming (ACLP) can be used for sound analysis of policy systems
expressed as logic programs. However, the
completeness of policy analysis through logic programming is only guaranteed 
under certain conditions. Policy analysis also relies upon the closed world 
assumption, preventing the analysis from showing a property is satisfied for 
every finite domain.

There have been several proposals for applying logic programming to
distributed computing. This talk advocated for a distributed state machine
model of distributed computation. Datalog plus a notion of time can be used
to clearly represent states and state transitions. The specification of
the distributed computation is fully declarative on this model allowing direct
application of logic-based analysis to the distributed program. Logic
programming languages for distributed computations are expressive. Operational
properties can be represented in the language. Both synchronous and
asynchronous communication models can be represented.
However, scalability remains an open problem.

\mypar{Concurrent and Distributed Systems, Integration --- Panel discussion with panelists Manuel Hermenegildo, Boon Thau Loo, Teri Swift, and Jan Wielemaker, moderated by David Warren}

Warren asks, ``What is the biggest current stumbling block in your
applications?'' The question is addressed first to Swift.
\begin{description}

\item {Swift}'s response is based on her work as a consultant for customs.
She points out that Prolog requires many extensions to produce useful 
applications.  In this case, it was extended for use as a server and for elastic search using Java, and
use of Prolog is restricted to data standardization. She laments that she ended up
with a Java program that included a little bit of Prolog.

\item {Loo}'s response concerns the use of Prolog for declarative 
networking. He points out that we face problems with usability, and with 
integration with legacy systems. In addition, he states that there is a problem
getting people to adopt declarative networking using logic programming. 

\item {Wielemaker} states that SWI-Prolog has been designed to work well with
legacy systems and other systems and languages. He is currently working on a 
program in which Java components are being gradually replaced with Prolog 
components.

\item {Hermenegildo} points to the continuing scalability 
problems for logic programming applications. In addition, there is a significant
adoption problem regarding newer, more advanced features that address these 
shortcomings.

\item Swift states that SWI-Prolog is the closest to a usably integrated logic 
programming language, but is still not quite there. She wants to integrate 
cluster computing as well.

\end{description}
Warren asks, ``In a multi-language, multi-software ecosystem, what are the
roles of logic programming languages?'' The question is first addressed to 
Wielemaker
\begin{description}

\item Wielemaker responds that there is no good general answer to this question.
It is necessary to look at each language and compare it to the features offered
by Prolog to determine how they can be effectively integrated.

\item Hermenegildo says that for some uses, for example provers, some LP 
software integration is automatic. But, there are still cases, such as 
networking, where careful deliberation is required to make a good choice about
how to integrate LP.

\item Loo adds that, regarding network protocols and domain-specific languages, policy evaluation is
the right use of LP. By contrast, data-level applications require a low-level
approach.

\item Swift contends that we should reconsider the distinction between 
low-level and high-level problems/applications/programming. She says that 
Scala is an example of a language that does both low-level and high-level well.

\item Wielemaker proposes that Prolog is more flexible than it may seem. It can
manage many different data models, but appropriate interfaces must be provided.

\end{description}
Warren asks, ``What should the underlying LP technology you use do better
to support your networking applications?'' The question is first addressed to
Loo.
\begin{description}

\item Loo answers that there is a persistent problem with
distribution in Datalog and Prolog. It is necessary to make the computations 
asynchronous. 
As a result, actual protocol implementations 
require extensions to Datalog and user-defined functions to work correctly. 
These extended programs become messy. It is no longer possible to prove their 
correctness without oversimplifying them.

\item Swift argues that there are already too many specialized features in XSB.
They do not work together well, and it takes a lot of time to figure out which
special features to use and how to integrate them together.

\item Wielemaker explains that the SWI-Prolog approach is to cover as many 
features of LP as possible, without doing any of them exceptionally well. He 
tries to do everything at a sufficient level of quality.

\item Hermenegildo agrees that special systems require specialized 
technologies. LP needs a good architecture for loading and unloading relevant, 
specialized features. He thinks we need standardization.

\item Loo concurs that standardization would be a good improvement. He thinks 
that LP should follow the software engineering model and offer an open API.

\end{description}
Warren asks, ``Is there a practical role for parallelism in your existing
applications?''  The question is first addressed to Hermenegildo.
\begin{description}

\item Hemenegildo answers, ``Yes, all the time''. We have the hardware and we 
want to use it for LP applications. Funding for parallelism was solid in the 80's and 90's, but disappeared 
because the predicted bar to faster processors did not happen. 
But without that support, and now with the need for parallelism that can take advantage of the ubiquitous multi-core processors, the
current technology for parallelism is not as good as the technology that was developed 
during that period of interest in parallelism.

\item Wielemaker also answers yes. He states that threading is good enough for
many tasks, but for interdependent tasks or parallel computation of a single
task, multi-threading is not good enough.

\item According to Loo, Parallel Datalog can be used for low-level networking 
tasks. One needs to pay special attention to the ordering. Ordering constraints make
the problems more interesting.

\item Swift contends that shared-memory parallelism is not as important as
distributed parallelism. There is no need to reinvent the wheel here. There are
many teams working on parallel computation. We should integrate their results
into our work on LP.

\item Hermenegildo says that we should try simple, intermediate approaches.

\end{description}

\mypar{Applying ASP in Industrial Contexts: Lessons Learned and Current
Directions --- Invited talk by Francesco Ricca}

This talk offered reflections on experiences using Answer Set
Programming (ASP) for industrial applications. ASP is a declarative programming
paradigm for logic programming and non-monotonic reasoning using stable-model
semantics. There are several robust and efficient implementations, including 
DLV, Wasp, Clasp, CModels, and IDP. Areas of applications include 
artificial intelligence, knowledge representation and reasoning, information 
integration, data cleaning, bioinformatics, robotics, etc.

The talk describes the use of DLV and Wasp to build 
industrial applications in many of these areas and gives sustained attention to
two of these applications. First was the use of DLV to build a system, ZLog, that
classifies customers of a call center and routes the customer's call to the most
appropriate service. Categories are created by experts (call center operators)
using a friendly GUI in ZLog, and then automatically translated into ASP rules.
Category definitions and the customer database are fed to DLV, which computes
the class of customers that fall into each category. Category classification 
then determines routing of call. ZLog runs in a production system at Telecom 
Italia where it handles 400 calls per second (over one million calls per day), 
and classifies each customer calling in in less than 100 ms.

The second application is a Team Builder for scheduling teams
of employees at the Gioia Tauro seaport. Scheduling is subject to many 
constraints: number of workers of each role necessary for a shift, contractual 
constraints on employee assignments, fair distribution of workload and 
assignment to heavy or dangerous roles, etc.  Manual team
building took hours, and mistakes are costly in terms of team performance and
penalties for contractual violations. The Team Builder application built using 
ASP offers a friendly GUI and guarantees that all constraints will be respected.
For the seaport, Team Builder manages 130 employees and fills 36 meta-plans per
week. Shift assignments for a day can be done in seconds and shift assignments
for an entire month can be computed in less than ten minutes.

Several lessons emerged from the experience
developing industrial applications using ASP. ASP can be effectively applied to
real-world, industrial-scale problems. ASP applications can be rapidly
developed, and are easy to understand, maintain, and extend because ASP is a
purely declarative language. In order to realize these benefits,
ASP programming requires the support of development tools, like IDEs, and
integration with other programming languages and established development
processes and platforms. These are necessary to indicate that ASP is not just
for researchers. Such tools (ASPIDE and 
JASP) are being built, but more work is required. Also the ASP solver WASP has
been extended to enable handling of problems, such as the Partner Units Problem
and the Combined Constraints Problem that state-of-the-art ASP solvers cannot
manage.

\mypar{Building Large-Scale, Knowledge-Based Systems with ASP ---
Presentation by Gopal Gupta}

This talk described several challenges that arise for 
building large-scale knowledge representation and reasoning systems using ASP,
and then proposed a possible solution. Most ASP systems
rely on SAT solvers, and this approach raises several issues for large-scale, 
knowledge-based systems. First, the program has to be finitely groundable,
preventing the use of complex data structures to organize a large knowledge
base. Second, grounding the program can result in exponential blowup, which is
not feasible for a large knowledge base. SAT solvers will find the entire model
of the program, which may contain a lot of unnecessary information, and
hide the answer we are looking for in the model. Finally, minor inconsistencies
in the knowledge base will prevent the system from finding an answer set, but we
cannot expect large knowledge bases to be entirely free of inconsistencies.

The proposed solution is a query-driven ASP system that
supports the use of predicates. By making the system query-driven, the system 
searches only the part of the knowledge base relevant to the query and produces
only a partial answer set. The query-driven approach addresses the concerns
about unnecessary information in the answer set obscuring the answer we actually
want. By supporting predicates, the system can execute programs without
grounding them first, which addresses the issues regarding the use of data
structures and the exponential blowup of the program. The final issue is
addressed through incremental consistency checks as the relevant parts of the
knowledge base are explored. If inconsistencies do not exist in the relevant
part of the knowledge base, then they can be ignored.

This approach has been implemented in the s(ASP) language.
s(ASP) is an extension of Prolog with stable-model semantics, allowance of
general predicates, and goal-directed, query-driven execution.
Several applications have been written using this language but challenges
related to efficiency persist.

\mypar{Knowledge and Constraint Systems, Integration ---
Panel discussion with panelists Marc Denecker, Michael Gelfond (could not attend), Torsten Schaub,
Mirek Truszczynski, moderated by David Warren}

{Warren} asks ``In some of your applications, are there issues of 
scaling to very large data sizes?'' Question was first addressed to Schaub.
\begin{description}

\item {Schaub} responds Yes, and that data mining is the application
where this problem is most apparent. We must treat problematic predicates
specially by outsourcing them to a dedicated propagator. Logistics and robotics
both require the generation of many new constants for discovered objects. The
problem is not as bad for bioinformatics. Schaub is working on a new system that
translates ASP into SAT and scales to 4000 objects per domain.


\item {Denecker} argues that some scale problems are not caused by the combinatorics
of the search, but by the size of the knowledge that needs to be explicitly represented. 
I.e., the grounded programs are too large. This is a technical problem. A possible solution is
lazy grounding.

\item {Truszczynski} contends that scalability problems are unavoidable for
hard problems. The only genuine solutions are luck, in the form of patterns in 
the data, or excellent heuristics.

\end{description}
Warren asks ``Do we have to compromise on pure declarative programming to get 
programs to run efficiently?''
\begin{description}

\item Schaub responds that in Clingo, heuristics can be specified at the ASP 
level.

\item Denecker says that practical applications require whatever is necessary
to make them work. Systems that use the same knowledge base to solve many
different kinds of problems cannot be tainted with procedural tweaks.

\item Truszczynski states that it may not be realistic to expect actual systems
to be that pure.

\end{description}
Warren asks ``What are the limitations of the LP technology you use in your
applications?''
\begin{description}

\item Schaub responds that LP is no longer viewed as proof.  People are using ASP for specifying 
constraints, but it is hard to model reactions in ASP as constraints.  One needs
linear equations for that.

\item {Gopal Gupta:} One needs goal-directed implementations.

\end{description}
Warren asks, ``What are, and what should be, the roles of 
declarativity and procedurality in your applications?'' Question was first addressed to
Denecker.
\begin{description}

\item Denecker responds that one example is interactive search applications. 
Heuristics do not matter in this context, because they are replaced by user
choice. Such applications require sticking to a declarative approach.

In general we want systems in which
interfaces are built or composed of procedural code that will interact with a
declarative knowledge base and solver.

\item Truszczynski contends that it is critical that the knowledge base be
represented declaratively. The role of procedural programming is for use in 
individual reasoning tasks performed on the knowledge base. A declarative approach should be 
used to describe knowledge.

\item Schaub expresses disagreement on this point. He argues that control must 
be exerted over the reasoning process. He thinks this can be solved by using 
procedural languages to build interfaces. This is necessary for 
number-crunching, but procedural language may not be necessary to control the process 
reasoning---this can be done declaratively.

\end{description}
Warren asks, ``What are good new applications for logic-based 
technology?'' Question was first addressed to Truszczynski.
\begin{description}

\item Truszczynski replies that bioinformatics is one such application, where 
good work is being done by Schaub's group. 
Something must be done for network security, such as anomalous behavior 
detection or policy specification. There are several other areas:
declarative network management; decision theory, for modeling agent preferences 
and resolving conflicts, where research problems are in need of better support for optimization
(Schaub's Asprin system that incorporates preferences into the ASP solver may help); and program derivation, for 
generating a program from a natural language specification of the problem.

\item Schaub adds that the problem for bioinformatics is that we cannot model 
non-linear constraints. He also mentions logisitics, as well as situations
in which many different reasoning tasks are required to solve a problem, for
example cognitive robotics.

\item Denecker says it is a problem of conquering infinite space. Some applications
need to go from view to data.

\item Truszczynski adds that representation of, and reasoning about, uncertainty
is critical.

\end{description}

\mypar{Declarative Probabilistic Programming --- Invited talk by C. R. Ramakrishnan}

This talk presented an extension of logic programming to support 
probabilistic facts and probabilistic reasoning, and described the current state of probabilistic programming
languages and their applications.

Logic programming is good for providing an executable 
specification of operational semantics.  For example, a simple set 
of rules captures the semantics of the lambda calculus. Logic programming 
can do the same for abstract semantics. Context-insensitive pointer analysis can
be specified with a small number of logic programming rules that look just like
the formal inference rules used to define the pointer analysis.

Logic programming can also be used to build a model
checker for logics with temporal properties, as exemplified with 
Computational Tree Logic (CTL). CTL contains formulas that describe states of 
systems and the formulas that describe paths (sequences of states). 
Tabled resolution is required for the model checker to terminate,
but query evaluation will be dynamically stratified. The time complexity for 
model checking is $O(|T| \cdot |\varphi|)$, where $|T|$ is the total number of
states and transitions and $|\varphi|$ is the size of the formula $\varphi$. The
space complexity is $O(|S| \cdot |\varphi|)$, where $|S|$ is the total number of 
states.

Probabilistic logic programming languages
combine rules with probabilistic facts. The result of the query is a 
probability distribution. There are several systems for probabilistic logic 
programming, including ICL, PRISM, ProbLog, etc. 
Probabilistic logic programming can be applied to build model checkers for systems that
have probabilistic temporal properties. Tabling is needed for
probabilistic model checking, just as it was for non-probabilistic model 
checking.  Inference for probabilistic temporal logics requires
additional support to track and differentiate random variable valuations in
different system runs and to ensure the computation terminates even though there
may be infinitely many distinct system behaviors.

The performance of probabilistic LP-based model
checking is comparable with other model checkers. It provides the first
realistic model checkers for expressive languages, such as pi-calculus, mobile
ad-hoc networks, multi-agent systems, etc. 
It supports the first implementation of a model
checker for the GPL language. Ramakrishnan also introduced XPL, a logic for 
reasoning about systems that feature both probabilistic and non-deterministic
choice.

\mypar{Horn Clause Solvers for Network Verification --- Presentation by Nikolaj Bj{\o}rner}

This talk discussed the results of attempts to represent networks
as Datalog programs, and then to use Horn clause solvers in Z3 to perform
verification of properties of the network. Z3 has been used to find solutions
to symbolic representations of problems from many domains. It has been applied
to program analysis to determine whether an execution path is feasible (SAGE),
whether policies satisfy a given contract (SecGuru), etc. In addition, Z3
contains several specialized engines when the formulas to which it is applied
are constrained Horn clauses.

Horn clause solving is useful because networks can be expressed as 
Datalog programs using constrained Horn clauses, in which packets are 
represented as differences of cubes.
For example, entries in a routing table can
be expressed by rules where a predicate representing the current location
of a packet and its source and destination is in the body of the rule, and
the next hop location is in the head of the rule. The body will also contain
constraints concerning the range of addresses that determine the next hop from
the current location of the packet. Given this representation of the network,
the problem of computing all the packets that will reach a given destination
from a given source becomes a reachability problem that Horn clause solvers
handle well.

The experience so far is that Z3 provides a general interface for
network verification problems represented as Horn clauses, but that the Horn
clause engines within Z3 each work on a very select set of problems.

\mypar{Logic Programming from NLP to NLU? --- Presentation by Paul Tarau}

This talk describes the application of logic programming to Natual Language
Understanding (NLU).  One of the original motivations for
logic programming was Natural Language Processing (NLP). There is
an opportunity now for logic programming to help us achieve some of the
objectives of NLU. Currently, statistical approaches to NLP are prevalent
(through machine learning and ``deep learning''), but for NLU we want observable
behavior and human understandable output. We can accomplish this by using some
form of logical representation, which is the closest formal 
mechanism to natural language.

There is a rich assortment of both logical forms and logic
programming tools available for building NLU systems. Graph-based NLP
algorithms, such as TextRank, are effective, but do not feature NLU elements.
They could be extended to NLU by using richer graphs built with logical
representations of sentences. Another application, developed by the presenter and his
colleagues, builds natural language-enabled agents using Prolog.  Beyond Prolog,
both constraint programming and ASP systems can provide
LP-based support for the transition from NLP to NLU. Tarau mentions several
applications for NLU that are all emerging now, such as interactive story
telling, voice-enabled software agents (e.g., Siri, Cortana, etc.), home
automation systems and IoT, and even search engines are moving toward becoming
NLU-enabled question answering systems.

\mypar{Directions for Research and Applications: Big Data
Analysis in Depth and Scale --- Open discussion moderated by David Warren}

{Warren} asks the general audience, ``What are the most important, real-world logic 
programming applications?''
\begin{description}

\item {Jeffrey Rosenwald}: it is transforming unstructured logging
data into something that a human can look at, or log analysis that can reveal 
anomalous data.

\item {Paul Tarau}: it is Prolog-based Natural Language Processing
for interactive agents for games and story telling.

\item {Gopal Gupta}: it is the Internet of Things, processing states and
observations for a huge numbers of sensors.

\item {Torsten Schaub}:it is assisted-living applications for smart
homes. Careful research is being displaced by ad hoc solutions built into
gadgets that are already being produced.

\end{description}
Warren asks, ``What is the role of logic programming in Big Data?''
\begin{description}

\item {Jan Wielemaker} answers that it is intrusion detection in networks. More
generally, logic programming can be useful for preprocessing data from
high-volume sources.



\item A participant says that he tried machine learning (supervised learning) for classification
on Big Data, and explains that we can use logic programming rules to handle
cases which machine learning classifies as ``unknown'', which is a much smaller
data set.

\item Rosenwald says that his system for materials handling and logistics collects gigabytes of data everyday, and logic programming can be used for filtering to
reduce data size.

\item {Tuncay Tekle} states that LogicBlox is using Datalog to process Big Data
for retail solutions.  To prove the effectiveness of 
logic programming to the skeptical, we need to show that the additional revenue
generated by LP solutions outweighs the cost of implementing and deploying the
LP solution.


\item {Annie Liu} closed by suggesting that we should build a repository of logic
programming applications.

\end{description}

\pagebreak


\mytitle{Solver-Aided Declarative Programming \myabs}
\myauthor{Molham Aref, LogicBlox, Inc.}

\begin{abstract}
I will summarize our work on a declarative programming language that offers native language support for expressing predictive (e.g. machine learning) and prescriptive (e.g. combinatorial optimization) analytics. The presentation gives an overview of the platform and the language. In particular, it focuses on the important role of integrity constraints, which are used not only for maintaining data integrity, but also, for example, for the specification of complex optimization problems and probabilistic programming.
\end{abstract}



\mytitle{Logic Programming in the \\ 
  Materials Handling and Logistics Industries\\}

\myauthor{Jeffrey A. Rosenwald, 
Intelligrated, LLC}

For quite some time, I have been using SWI Prolog to build systems that are used to
control machinery that moves packages around 1 million sq.\ ft.\ warehouses. These
kinds of systems are deployed in large retail distribution centers, airports, and
parcel and postal sortation centers.

\mysec{Characteristics of the application}
\begin{itemize}

\item {\bf Systems are big}---it is not unusual for a one sorter to have 20,000 alarm
points.
\item {\bf High throughput}---range is typically 10\textsuperscript{5}  to 10\textsuperscript{6}  items sorted per day
\item {\bf High accuracy}---sortation mistakes are expensive. Error rate is about 1 in 10\textsuperscript{4}
items sorted.
\item {\bf Heterogeneous}---different systems, platforms, and vendors may be involved,
\item {\bf Event-driven}---everyone reacts to events that occur on the machinery
\item {\bf Soft real-time}---end-to-end service times are measured in milliseconds.
Some variability of service time is tolerable, within limits .
\item {\bf Fault-tolerant}---non-Byzantine failure, that is late or lost (infinitely late)
messages, are recoverable.
\item {\bf High availability}---many sorters run 20 hours per day, 364 days a year. The
execution epoch of the control system is measured in years. It is never taken
down in the absence of failure.
\item {\bf Reliable and durable}---the machinery is built to last for 25 years (many are
older than that)
\end{itemize}
\vspace{-1ex}
\mysec{Characteristics of the software design}
\begin{itemize}
\item An agent-oriented design provides a collection of small stand-alone programs
---a community of Holons organized in a Holarchy.
\item Each Holon has a sphere of influence and a protocol.
\item Holons communicate with one another anonymously by message passing
using a publish and subscribe regime that spans the entire CPU cluster.
\item The Holarchy can be deployed across a cluster of several CPUs, which provides
for hot-standby redundancy, load sharing, and automatic fail-over.
\item Holons in the system play the following roles:
  \begin{itemize}
\item
two Holons interact with Siemens S7 Programmable Logic Controllers
(PLCs) via TCP/IP byte streams to provide system control, alarm, and
diagnostic surveillance,
\item one Holon is responsible for Routing. That is, where an item ought to go
when it's seen (by a bar code scanner) at a particular place and time on the
machinery,
\item several Holons provide GUI elements for various control and status
displays.
  \end{itemize}
\end{itemize}
\vspace{-1ex}
\mysec{Characteristics of the implementation}
\begin{itemize}
\item The system is fast, small, flexible, scalable, resilient, easy to understand and
maintain, and darn-near bullet-proof.
\item The amount of source code  is about 20K lines of Prolog. This is
  nearly {\bf an order of magnitude smaller} than it's equivalent Java based counterpart
\item IPC messages are human-readable.
\item The message pattern is loosely-coupled.
\item Holons have almost no code dedicated to debugging.
\item Debugging of dynamic behavior is done by lurking/logging on the broadcast
channel.
\item Unit testing of Holons is a highly effective method of eliminating defects.
\item Badly formed messages cannot enter the system. They are detected early and
discarded.
\item Holons can be replaced on-the-fly, without taking the system down.
\item The design is naturally parallel. Effective use of multi-core CPUs is free.
\item Association, Parsing, and Pattern Matching are essential.
\item Non-deterministic Content Addressable Memory that is afforded by the
language provides an elegant solution to many problems.
\item Base system does not have a relational database.
\item Voluntarism is an important concept: Holons do things because it is in their
nature to do so. They do things because they want to, not because they have
to.
\end{itemize}
\vspace{-1ex}
\mysec{Barriers to adoption}
\begin{itemize}
\item Materials Handling is an old business with many large entrenched players,
with vested interests
\item Selection of technological platforms and systems at this scale is a business
decision, not a technical one.
\item Fear, Uncertainty, and Doubt
\item Market penetration will likely be revolutionary, not evolutionary
\item Irrational fear of Open Source generally and of the GPL in particular
\item Incompatibility of the GPL with business models that espouse ``ownership'' of
intellectual property.
\item Prolog is almost unknown in the U.S., outside of academic circles.
\item Angel Investors and Venture Capitalists are reluctant to invest in ideas that do
not produce sustainable competitive advantage that can be sold. This usually
means intellectual property: patents, copyrights, or trade secrets.
\end{itemize}
Here's a video of a system deployed by Buemer/Crisplant (our European
sister company).  I would control the cross�belt sorter and induction
conveyor (between 00:50---02:10).
\url{https://www.youtube.com/watch?v=tiW7CSs9G8M}

\pagebreak

\def\bibdir{carvesato}


\mytitle{Concurrent Logic Programming: Met and Unmet Promises%
}%

\myauthor{
  Iliano Cervesato
, Carnegie Mellon University
\\
Edmund S.L. Lam
, University of Colorado, Boulder
}




\begin{abstract}
  Logic programming has been heralded as the quintessential
  declarative programming\linebreak[4] paradigm, although many
  instances provide extra-logical constructs that undermine this
  aspiration.  The word ``declarative'' conjures two promises: the
  first is the ability to write code that reflects a natural,
  human-friendly, description of the problem at hands as opposed to a
  mechanistic, hardware-oriented, encoding of a solution.  The second
  is the opportunity to reason logically about it, thereby
  automatically strengthening assurance, security and performance.
  While the first promise has been fairly successful in some domains,
  the second still has to live to its expectations.  We explore both
  promises in the context of concurrent logic programming.  We
  highlight them using CoMingle, a logic programming language designed
  to develop mobile Android applications.
\end{abstract}

\mysec{Logical specification of concurrent applications}

Concurrent and distributed applications have traditionally been
developed by writing a separate piece of code for each participating
device (or class of devices).  This \emph{node-centric} approach puts
the onus of handling messaging and synchronization on the programmer.
This is no small burden: on the messaging side, the programmer needs
to make sure that each sent message has a recipient and vice versa,
and that sender and receiver agree on its format --- simple tasks that
quickly become a time sink as an application grows larger.  The
synchronization side is more tricky as the programmer is left alone
battling the many pitfalls of concurrency (deadlocks, live locks and
unwanted race conditions) --- complex tasks even for small
applications.  Because the code running on each device is a separate
control flow, little automation is available to alleviate these
concerns as current program analysis techniques typically focus on
individual control-flows and do not work well for reasoning about the
executions of a concurrent program as a whole.  These effects are
compounded by the fact that, as mobile applications become
commonplace, many of them are being developed by programmers with
relatively little training or experience.

An alternative approach is to write a unified program that captures
the behavior of a distributed application as a single entity.  This
\emph{system-centric} approach gives the programmer a bird-eye's view
of the behavior he/she is trying to achieve.  Being a single program,
it is easier to automate basic checks such as message format
consistency, for example as a form of type-checking.  This unitary
system-centric specification is automatically transformed into the
node-centric code that runs on actual devices through a process called
choreographic compilation.  It is this transformation, rather than the
programmer, that handles the tedium of managing communication and the
intricacies of getting synchronization right.

While the system-centric approach to programming distributed
application is not exclusive to the paradigm of logic programming (in
fact, it underlies many of Google's applications~\cite{GWT}), logic
programming is proving particularly well-suited for this
purpose~\cite{Ashley09iclp,Grumbach10padl,Lam15coordination,Loo06sigmod}.
Logic programming provides a natural way to write specifications that
represent how the distributed computation proceeds as a whole rather
than forcing the point of view of any specific node.  One language
that embraces this philosophy is CoMingle~\cite{Lam15coordination}.
CoMingle is a rule-based language for programming mobile distributed
applications, originally Android apps.  CoMingle implements a fragment
of first-order linear logic using a forward-chaining semantics, as
found in languages based on multiset-rewriting such as
CHR~\cite{Fruewirth09book}.  It enriches it with sorts (making it a
strongly-typed polymorphic language), locations (which identify
computing nodes), and multiset comprehensions (which provide a natural
mechanism to manipulate arbitrarily many facts matching a given
pattern).  Specifically designated atomic facts allow CoMingle to
trigger local computations and respond to them (used for example to
process input from an Android device or to render output on the
screen).  We used an advanced prototype of CoMingle~\cite{comingle} to
implement a number of mobile applications.  We were able to write each
of them in a few hours, which compares favorably with the standard
node-centric approach.  We built one such application both using
CoMingle and by writing traditional code: the former was about one
tenth of the size of the latter with no noticeable difference in
performance~\cite{lam15tr}.  This ease of development gave us time to
experiment with application-level features, with new communication
behaviors typically taking minutes to implement.


\mysec{Reasoning about concurrent applications}

Because in its purest form a program is a logical formula, logic
programming has often been trumpeted as facilitating reasoning about
one's code, where reasoning is variedly understood as providing
provable assurances of correctness, guaranteed performance, and more
recently security.  With a few exceptions
(e.g.,~\cite{mcallester02jacm} about performance bounds), we argue
that such expectations of correctness have not been met.  For example,
correctness presupposes a specification that can be compared with an
implementation, but rarely does a programmer write two such formulas
for the same problem, and in any case tools to verify the expected
subsumption are rarely available.

Concurrent logic programs, for example the ones we wrote in CoMingle,
similarly come short of availing themselves of the reasoning
possibilities of the underlying logic.  The consequences are somewhat
more dire in this setting as writing concurrent programs is much
harder than developing code that does not engage in synchronization.
The proliferation and ease of deployment of mobile apps, again often
developed by novices, means that there is a lot of buggy code out
there, with much more to come.

Even in a large program, a fairly small part of the code of a
distributed application is about concurrent interactions, often with
recurring patterns (this is particularly evident in CoMingle programs,
where inter-node communication and local computation are written in
separate languages --- CoMingle itself and Java, respectively).  We
postulate that this is an opportunity for logic-based methods, if not
wholesale logical reasoning, to participate in the development of
concurrent and distributed applications in the form of formal analysis
tools.  One promising idea is session types~\cite{honda93concur},
which describe the communication pattern of a program, thereby
allowing the implementation of a tool to statically catch messaging
errors and deadlocks.  Session types are currently limited to
relatively simple interactions, but they are rapidly being developed
to handle larger classes.  Other techniques include logic-based
modularity~\cite{cervesato15lpar}, which gives the programmer control
over the scope of interactions (in contrast to the traditionally flat
name-space of logic programming).  One last class of techniques that
holds substantial promises in the development of correct concurrent
programs specifically is coinductive reasoning, for example in the
form of bisimulation.  While tools are still in their infancy, the
growing realization that many program properties are coinductive in
nature are sure to accelerate their development.

What logic programming does is to give the programmer a language that
abstracts some idiosyncrasies of the underlying machine, ultimately
letting him/her write less code: abstraction, not reasoning, makes
small programs easier to get right.  But as programs grow, coding
complexity creeps back up, with little available to the programmer to
manage it in a typical logic programming language.  We postulate that
the largely untapped reasoning potential of logic programming in
general, and concurrent logic programming in particular, holds the
promise provide assurances that is elusive in other paradigms.




\mytitle{What Tweety-the-Penguin and Faulty Suitcases Tell Us \\about Productivity, Cybersecurity and Data Sciences \myabs\\}
\myauthor{Marcello Balduccini, Drexel University}

\begin{abstract}
The areas of research of commonsense, reasoning about actions and change, and constraint satisfaction have a long-standing tradition in the knowledge representation community. These areas have frequently developed independently of each other, but various forms of their combination have proven extremely useful for practical applications.

In this talk, we aim to convey some sense of the breadth of applications
yielded by the research at the intersection of commonsense, reasoning about
actions and change, and constraint satisfaction. We will start from our
early, and somewhat unexpected, success in solving industrial-sized
problems with a planning and diagnostic system for the Space Shuttle, and
we will then expand to later work on hybrid reasoning, industrial
scheduling, cybersecurity, and information retrieval.  
\end{abstract}

\pagebreak

\def\bibdir{russo}


\mytitle{
Distributed Systems Management:\\
Logic Programming Solutions and Challenges\\
}

\myauthor{
Jorge Lobo
, ICREA \& Universitat Pompeu Fabra
\\
Alessandra Russo 
and 
Emil Lupu
, 
Imperial College London
}






Within the last decade, distributed systems have rapidly evolved from applications that run within local networks, using  simple client/server architecture, to applications that run over complex large-scale networks and large-scale platforms across multiple administrative domains and geographical areas. This trend is set to continue with the increased deployment of embedded devices and Internet of Things technologies. Configuring and managing such systems is a significant challenge due to their openness and extensibility, their dynamic nature - new components appear, disconnect or migrate to new locations, and the many different functions for which management is required - failure, security, performance, accounting etc. To address these challenges rigorous and scalable solutions are required, which are able to adapt to the dynamic changes in the system and in the environment in which they operate.


For many years we have been using tools and techniques from AI for tackling management issues of distributed systems and networks working towards the aim of building autonomic management systems and we have seldom found ourselves proposing solutions in which logic programming plays a central role. 
Below we describe some of the technical challenges underpinning the development of solutions for configuration and management of distributed systems in the context of three different application domains -- policy, security and distributed system management -- and highlight the advantages that our LP approaches have provided to overcome them. We also briefly describe open issues and suggestions for future research that would lead these results to real industrial applications.



\mysec{Policies in system management}

Policy is a very generic term. In the context of system management, policies define how choices that affect the behaviour of systems should be governed.  The aim of policy-based approaches to systems management is to provide a separation of the rules that govern the behaviour of a system from the actual functionality provided by that system~\cite{Sloman1994}.  Policy can describe how to handle failure, security, performance, accounting, etc.  
In our work we have developed dialects of LP for specifying and implementing policy (e.g.~\cite{Lobo1999}), for automated policy analysis (e.g.~\cite{Russo2002},\cite{Bandara2003}), conflict resolution (e.g.,~\cite{Jobo2003}) and refinement (e.g.,~\cite{Bandara2004},\cite{Craven2009}). The formalisation of action and change in LP has had tremendous impact in our work. 

 The earliest work is the {\em event-condition-action} {\cal PDL} introduced in~\cite{Lobo1999}.  Its declarative semantics is founded on formal descriptions of action theories based on automata (e.g., \cite{Baral1997}). The key challenges were the development of a language that can be (efficiently) implemented, that provides concise representation of policies, and that has a formal (declarative) semantics, a crucial aspect in order to enable automated analysis and conflict detection. In~\cite{Bandara2003}, an Event-Calculus language has been proposed to model management policies including authorization policies and uses early results in abductive LP for the analysis of event-based specifications~\cite{Russo2002} to automatically detect modality conflicts and application specific conflicts in the policies.  With abduction not only policy conflicts are detected but also explanations are generated on how these conflicts may arise. 
The use of LP and abductive-based solution has helped us formalising the policies in a rigorous manner and consequently reason about their correctness. On the other hand, the challenge presented by dynamic aspects of distributed systems managed by these policies could be hinder by the close world semantic or the notion of existing (fix) background knowledge of an abductive framework. We were able to overcome these limitation by developing a novel multi-threaded distributed abductive algorithm and system  \cite{Ma2008} that is {\em open} in that it can opportunistically make use of dynamic changes of the system configuration during the reasoning process.

One of the most difficult questions in policy management is how to do refinement. In system management, policies are related to high-level system goals (requirements), be these functional or not functional. The challenging task is how to support automated refinement (or generation) of operational policies (e.g., policies that the system is able to enforce) from high-level system requirements. In \cite{Bandara2004} we have defined policy refinement as a process of realisation of requirements expressed in terms of high-level abstract entities into policies expressed in terms of concrete objects/devices that when performed will achieve the high-level goal.
This approach, however, assumes the existence of a complete LP representations of the domain (concrete objects/devices). Again in the context of dynamic systems, where components might ``ad-hoc'' join or leave the system (e.g. dynamic service composition), or high-level requirements might be changed at run-time, the refinement process would need also to be dynamic. So the key challenge for LP, in this case, would be how to support run-time refinement.

\mysec{Security}

Security (and privacy) is perhaps the area of system management most easily identified with policy-based models and in which LP as modeling tool has had the most impact (e.g. \cite{becker2010secpal},\cite{gurevich2008dkal}).  \cite{Bandara2003} already touches on issues related to access control in the context of general policies but the work in~\cite{Craven2009} addresses the challenge of specification and analysis of policy in which the enforcement requires monitoring the system over time. A typical example of these types of policy are obligation that an entity must fulfill in the future in order to obey policy.  An LP dialect is used to described policy and an abductive procedure tailored for the analysis is introduced.

In order to avoid conflicts and non determinism, many management systems use ordered sets of rules in their configuration. In such scenarios, the first "matching" rule is executed whilst the others are ignored; firewall rules being a classical example. Generating configurations that can be directly deployed into existing systems, i.e, without adding additional interpreters requires therefore to synthesise not only the rules but also their ordering, whilst preserving desired properties. We have successfully used argumentation in LP for this purpose to synthesise complete firewall configurations on real examples~\cite{DBLP:conf/im/BandaraKLR09}. 

Trust Management Systems (TMS) are concerned with distributed access control in which policies defining under what conditions a subject is able to access resources expect to get credentials from the subject prior to evaluation. Reciprocally, the subject may have policies that condition the server providers (of the resources) that the subject will accept. These policies may also need credentials from the server provider before evaluation. Little has been known about system independent formalizations of TM. Recently, a general axiomatization for logic-based trust management with a Kripke model theoretic semantics and a Hilbert-style proof system was introduced in~\cite{becker2012foundations}. Our contribution in~\cite{pasarella2015reasoning} was to link \cite{becker2012foundations} to results from Miller's scoping and modules for LP and provide an alternative axiomatization that allowed us to define and implement an ASP-based theorem prover, but most importantly, to establish the computational complexity of doing proofs in this logic.

\mysec{Declarative distributed computing}
More recently, we have directed our attention to a much broader class of problems in distributed systems and network management, with the aim of developing a general but rigorous framework for formalising distributed algorithms, analysing and reasoning about their correctness in situations where networks are dynamic (e.g. nodes can join and/or leave the network during the execution). These are hard and open problems that are not only relevant in network management but to any application domain that requires distributed computations.  Again, building upon theoretical results on reasoning about actions~\cite{Gelfond1998}, and their translations to causal logics, we have been able to propose a declarative approach to distributed computing called {\em D2C} (see \cite{Ma2013}). In {\em D2C} distributed algorithms can not only be specified as action theories of fluents and actions, but also executed as collections of (input/output) automata, and analysed using the results on connecting causal theories and Answer Set Programming (ASP). The declarative semantics has enabled us to provide automated translation of a distributed algorithm expressed in {\em D2C} and the underlying communication model (e.g. synchronous or asynchronous) from the causal logic-based specification into ASP programs, providing a framework for implementation and analysis. We have demonstrated the generality of our declarative approach by showing how it can be used to analyse different classes of network routing protocols, as well as execute distributed algorithms for pattern formation in multi-robot systems \cite{Ma2016}. 

\mysec{Directions of research}
\label{sec:openChallenges}

Much can be said about future research but we would like to conclude with just three questions that can be used as discussion points to think about future work:
\begin{itemize}
\item
Is policy learning an alternative to policy refinement? A better integration of numerical methods and logic programs will be helpful. 
\item
Is the perception that LP/Datalog is not efficient enough for high-throughput applications in security real?  
\item
What does it mean to do distributed logic programming? Is declarative networking the appropriate abstraction - implementations of distributed logic programs.
\end{itemize}

\noindent
Key problems common to the above three aspects of future research are the {\em scalability of LP} and the {\em close world} assumption of the domain. For example, in declarative distributed computing, the analysis of routing protocol, which is known to be an NP-hard problem, if based only on ASP computation can handle only toy networks composed of no more than four nodes. In our experience, it has sometime been the case that the time and space used for grounding a problem are too much to start considering real industrial application problems. The question is how can we improve the scalability and also how can we represent generic domains.

{

}

\def\bibdir{ricca}

\newcommand{\dlv}{{\small DLV}\xspace}
\newcommand{\aspide}{{\em ASPIDE}\xspace}
\newcommand{\jdlv}{J\dlv$\!$}
\mytitle{Applying ASP in Industrial Contexts: \\ Lessons Learned and
  Current Directions
 \\
}

\myauthor{Nicola Leone and 
Francesco Ricca, 
University of Calabria 
}


\begin{abstract}
Answer Set Programming (ASP) is a declarative programming para\-digm 
that has been proposed in the area of logic programming and
nonmonotonic reasoning.
ASP has become a popular choice for solving complex problems, as
witnessed by the numerous scientific applications that are based on
ASP, and it is
nowadays attracting increasing interest also beyond the scientific community.
We report on the development of some applications of ASP in industrial contexts.
We focus on the lessons we have learned and on current developments.
We outline the advantages of ASP from the software engineering point
of view, and
we stress the importance of extending tools and development
environments to speed-up and simplify the implementation of real-world
applications.

\end{abstract}

\mysec{Applying ASP in industrial contexts}
Answer Set Programming (ASP)~\cite{DBLP:journals/cacm/BrewkaET11,DBLP:journals/tods/EiterGM97,DBLP:journals/ngc/GelfondL91} is a declarative programming paradigm that has been developed in the field of logic programming and nonmonotonic reasoning.
ASP combines a high knowledge-modeling power~\cite{DBLP:conf/coco/DantsinEGV97} with a robust solving technology~\cite{DBLP:journals/ai/CalimeriGMR16}. 
As a consequence, ASP has become a popular choice for solving complex problems.
Indeed, it has been applied in numerous scientific applications in
the areas of Artificial Intelligence~\cite{DBLP:conf/lpnmr/BalducciniGWN01,DBLP:journals/tplp/GagglMRWW15,DBLP:conf/lpnmr/BaralU01,DBLP:journals/tplp/ErdemPSSU13}, 
Bioinformatics~\cite{DBLP:journals/tplp/ErdemO15,DBLP:journals/tplp/KoponenOJS15,DBLP:journals/jetai/CampeottoDP15}, 
and Databases~\cite{DBLP:journals/dke/MarileoB10,DBLP:journals/tplp/MannaRT15,DBLP:conf/ijcai/BravoB03,DBLP:journals/japll/BravoB05,DBLP:conf/sigmod/LeoneGILTEFFGRLLRKNS05}, to mention a few.
Nowadays, ASP is attracting increasing interest also beyond the scientific community~\cite{DBLP:conf/birthday/GrassoLMR11,InvitedFriedrich15}, and counts already some successful application in industrial products.
In particular, we report on our on-the-field experience in the development of some industrial applications of ASP, 
namely:\vspace*{-0.2cm}

\begin{itemize}
\item A platform employed by the call-centers of Italia Telecom, 
which classifies in real-time the incoming calls for optimal routing.
\item A tool for the automatic generation of the teams of employees~\cite{DBLP:journals/tplp/RiccaGAMLIL12} that has been employed in the sea port of Gioia Tauro for intelligent resource allocation.
\item A tool for travel agents for the intelligent allotment of touristic packages~\cite{DBLP:conf/rr/DodaroLNR15}.
\item An ASP-based platform for data cleaning~\cite{DBLP:conf/lpnmr/TerracinaML13} developed for analyzing and cleaning-up the archives of the Italian Healthcare System storing data on tumor diseases. 
\end{itemize}\vspace*{-0.2cm}

\noindent These applications were implemented by using \dlv~\cite{DBLP:journals/tocl/LeonePFEGPS06}, which is 
the first ASP system that is undergoing an industrial exploitation by a company, called DLVSYSTEM. 

\mysec{Lessons learned and current directions}

A lesson learned by developing real world applications is that ASP allows one to develop complex features at a lower (implementation) price than in traditional imperative languages. Indeed, the possibility of modifying complex reasoning task by editing text files, and testing it ``on-site'' together with the customer has been often a great advantage of the ASP-based development.
ASP can bring several advantages from a Software Engineering viewpoint, and the main qualities are flexibility, readability, extensibility, and ease of maintenance of ASP-based solutions.
Nonetheless, in order to boost the adoption of ASP 
in the scientific community and especially in industry,
it is important to provide programming tools 
that make easier the development of applications.
For this reason, we endowed \dlv with development tools 
conceived to ease the usage and the integration of ASP-based technologies
in the existing programming environments tailored for imperative/object-oriented languages.
In particular, we have developed two tools for developers: \aspide~\cite{DBLP:conf/lpnmr/FebbraroRR11} and \jdlv~\cite{DBLP:conf/kr/FebbraroLGR12}.
ASPIDE is an extensible integrated development environment for ASP,
which integrates powerful editing tools  with a collection of development tools for program testing and rewriting,  database access, solver execution configuration and output-handling.
JDLV is a plug-in for Eclipse, supporting a hybrid language 
that transparently enables the interaction between ASP and Java.

Currently we are working on application-driven improvements of ASP tools.
In particular, we are studying means for simplifying the extension of ASP systems with problem-specific heuristics so to speedup the evaluation of very hard real-world problem instances~\cite{ICLP16}.
Our experience in the development of systems and applications of ASP suggests that modifying an ASP implementation is very complex, and can be carried out effectively only by a few expert researchers. 
On the other hand, the implementation of a good heuristic requires a knowledge about the domain, which is likely to be found on people from industry.
Thus, we are working on the extension of the ASP system WASP~\cite{DBLP:conf/lpnmr/AlvianoDLR15} that allows one to easily plug-in, test and evaluate new domain-heuristics also by software developers working in the industry~\cite{ICLP16}.





\mytitle{
Building Large-scale Knowledge-based Systems with ASP 
}

\myauthor{Gopal Gupta, Kyle Marple, Elmer Salazar, Zhuo Chen, Arman Sobhi,
  and 
Sai Srirangapalli
\\
The University of Texas at Dallas
}


Answer Set Programming (ASP) \cite{baral} has emerged as a successful paradigm for developing intelligent
applications. ASP is based on adding {\it negation as failure} to logic programming under the stable model
semantics regime \cite{stable}. ASP allows for sophisticated reasoning mechanisms that are employed by humans
(common sense reasoning, default reasoning, counterfactual reasoning, abductive reasoning, etc.) to be
modeled elegantly. Numerous systems have been built to execute answer set programs that are
extremely sophisticated and efficient. CLASP is the best representative of these systems \cite{clasp}. 
These systems
restrict programs to predicates that only have variables and constants as arguments (general structures
are not allowed). Answer sets (or stable models) of such programs are computed by grounding the
program rules with the (finite) Herbrand universe, suitably transforming it, and then using a SAT solver
to compute models of the transformed program. These models of the transformed program are the
stable models of the original Answer Set Program. There are many problems with this model-finding
approach that rely on a SAT solver:

\begin{enumerate}

\item Since SAT solvers can only handle propositional programs, these approaches only work for
finitely-groundable programs. That is, programs with structures and lists occurring in arguments
of predicates cannot be executed, as grounding of such programs will result in an infinite-sized
program (due to the Herbrand universe being infinite). In many instances, lists and structures
are essential for representing information.

\item Grounding of the program can lead to an exponential blowup in program size. For programs to
be executable in such a system, a programmer has to be aware of how the grounding process
works and how the ASP solver works and then they have to write their code in such a way that
this blowup is minimized. This places undue burden on the programmer, as the programmer has
to have knowledge of the grounding procedure as well as the model-finding process.

\item If the number of constants in the program is large, then a SAT-based approach is infeasible due
to the size of the grounded program that will be created. It is next to impossible to build a
general-purpose knowledge-based system using such an approach, as such a knowledge-based
system will potentially have tens of thousands of constants.

\item SAT-based ASP solvers do not allow reasoning with real numbers.

\item SAT-based model-finding approaches compute the entire model. That is obviously an over kill.
Most of the time users are interested in a specific piece of information. Thus, if we have a
general purpose knowledge-based system, then the current ASP systems will compute the entire
model, i.e., everything that can be inferred from the knowledge-base will be computed.

\item Often, it is hard to isolate the solution that is embedded in the model that is produced by the
SAT solver. For example, if one solves the Tower of Hanoi problem using a SAT-based ASP solver,
then the answer set will contain a large set of moves that are in the model. One cannot easily
isolate the sequence of moves that represent the solution to the problem.

\item Since ASP systems compute the entire model, even a minor inconsistency in a narrow part of the
knowledgebase will result in the system concluding that no answer set exists. A practical, large,
real-world knowledgebase is very likely going to contain inconsistencies.

\end{enumerate}

We have been working on designing query-driven answer set programming systems \cite{goaldir}. A query-driven
system computes the partial answer set that contains the query (thus, it does not compute the entire
answer set). Having a query-driven system addresses problems 5, 6 and 7 mentioned above \cite{dcc},
however, issues mentioned in points 1, 2 3, and 4 above still remain as problems. To alleviate problems
1, 2, 3 and 4 above, we have extended our system to allow general-purpose predicates. Thus, our
extended system, called s(ASP), admits answer set programs containing predicates that are allowed to
have variables, constants and structures as arguments \cite{sasp-system,sasp-report}.

Our s(ASP) system does not ground the program. It can be thought of as full Prolog extended with
negation-as-failure under the stable model semantics regime \cite{sasp-report}. Problem 1, 2, 3, and 4 above are
eliminated by s(ASP), since programs do not have to be grounded prior to execution. The s(ASP) system
is publicly available \cite{sasp-system}, and has been used to develop a number of non-trivial 
applications based on ASP.
Some of these applications cannot be executed on traditional ASP systems such as CLASP, as these
applications make use of lists and structure to represent information. They have been developed by
people who are not experts in ASP. These applications include:

\begin{itemize}
\item A system for automatically performing degree audit of a student's undergraduate transcript at a
US University, i.e., automatically determining if a student can graduate with a degree or not.
The system represents the graduation requirements laid out in the course catalog as ASP clauses.
Use of negation is important for representing these requirements. The system has to make use
of lists, and has hundreds of courses that appear as constants in the program (hence its
grounding will produce an inordinately large program).

\item A system for disease management, particularly, for chronic heart failure. This system automates
the 80-page guidelines (that the American College of Cardiology has developed) by representing
them in ASP. While the current system can be run under systems such as CLASP due to the
number of constants not being too large, the final system that models a doctor's full knowledge
will have quite a few constants, and advanced data-structures may be needed.

\item A system that represents high-school level knowledge about cells (in the discipline of biology) as
answer set programs. It can answer high-school level questions posed as s(ASP) queries. The
goal is to represent the knowledge in the entire introductory biology textbook as an answer set
program, and then be able to automatically answer questions that would be asked of a student
(the questions have to be translated into ASP queries that are then executed to find the answer).

\item A recommendation system for birthday gifts: This system codes a human's knowledge about
friends, level of friendship, a person's wealth level, generosity level, and hobbies as answer set
programs. When queried, the system can recommend a birthday present for a particular friend.

\end{itemize}

We believe that the ASP paradigm is a very powerful paradigm that allows for complex human
thought processes to be elegantly emulated. Complex reasoning patterns that humans use can be 
elegantly modeled using ASP \cite{iclp16}. However, as argued above, the current model-finding, SAT-solver
based approaches are not able to realize the full-power of ASP. We argue that query-driven
implementations of predicate ASP are crucial to the paradigm's success. An additional advantage of 
a query-driven approach over model-finding approaches is that in the latter case, everything has 
to be modeled in the ASP paradigm, while in the former case both the standard logic programming
paradigm and the ASP paradigm can be made to work together.  

Significant progress has been
made with the realization of the s(ASP) system, however, considerable amount of research remains to
be done. We urge the community to invest effort in developing query-driven predicate 
ASP systems.

\medskip
\noindent
{\bf Acknowledgment:} Support from NSF Grant IIS 1423419 is gratefully acknowledged.


\mytitle{Declarative Probabilistic Programming for Program Analysis \myabs}
\myauthor{C.R. Ramakrishnan, Stony Brook University}

\begin{abstract}
Logic Programming has been successfully used for deriving efficient program analyzers and model checkers from succinct, high-level specifications. In this talk, we will examine what made logic programming especially suited for this task. We will survey some of the key technical developments that helped in these applications. We will also consider extensions to traditional logic programming semantics and inference techniques to treat probabilistic systems, and describe current work in the analysis of programs and models that use these extensions.  
\end{abstract}


\def\bibdir{bjorner}


\mytitle{Horn Clause Solvers for Network Verification}

\myauthor{%
Nikolaj Bj{\o}rner, 
Nuno P. Lopes, and 
Andrey Rybalchenko, 
Microsoft Research}


\begin{abstract}
We describe our experiences using solvers for Horn clauses 
with special emphasis on Network Verification. 
Z3 is a general purpose theorem prover with a plethora of special purpose
engines. Some of these engines are dedicated to solving Horn clauses. 
One use Horn clause solving is Symbolic Model Checking of software. 
Other uses are for checking reachability properties in packet switched networks.
Stratified Datalog can conveniently encode such properties, 
where the relations range over packet headers. Packet headers
are in turn bit-vectors. We developed Network Optimized 
Datalog (NoD) to solve Horn clauses originating from Network Verification.
\end{abstract}

\mysec{Horn Clause Engines in Z3}
This position paper highlights  specialized support for logic program analysis in
the context of Z3~\cite{z3}. Our main applications of Horn clause solving with Z3 are currently around 
software model checking and network verification. Z3 solves satisfiability of first-order logical
formulas modulo a set of built in theories, such as linear real and integer arithmetic, machine arithmetic 
(bit-vectors) and algebraic data-types. For the special case where formulas are \emph{Constrained Horn Clauses} (CHCs), Z3 admits
dedicated engines for solving satisfiability of these formulas. The available use cases are summarized in Table~\ref{tab:apps}.%
\footnote{The last three cases are un-tuned and are rarely used.}

\begin{table}
\centering
\begin{tabular}{|l|l|l|}
\hline
Application Area        & Solver                                       & To handle CHCs with \\ \hline \hline
Software Model Checking & SeaHorn~\cite{DBLP:conf/synasc/Gurfinkel15}  & Linear arithmetic, bit-vectors, arrays \\ \hline
Software Model Checking & Duality~\cite{DBLP:conf/cav/McMillan14}      & Linear arithmetic and arrays \\ \hline
Software Model Checking & PDR~\cite{pdr}                               & Linear arithmetic \\ \hline
Network Verification    & NoD~\cite{nod}                               & Bit-wise operations over bit-vectors \\ \hline
Points-to analysis 	& Finte Domains~\cite{DBLP:conf/cav/HoderBM11} & Bottom-up Datalog with hash-tables \\ \hline
Cyclic Induction proofs & Tabulation search                            & Algebraic data-types and arithmetic \\ \hline
Symbolic execution      & DFS SLD resolution                           & Quantifier free SMT constraints \\ \hline
Bounded Model checking  & Bounded BFS unfolding                        & Quantifier free SMT constraints \\ \hline
\hline
\end{tabular}
\caption{\label{tab:apps} Summary of Horn clause engines in Z3.}
\end{table}

By a CHC, we understand a formula of the form $\forall \vec{x} \ . \ \mathit{head}(\vec{x}) \leftarrow \mathit{body}(\vec{x})$,
where $\mathit{head}$ is either a predicate $p(\vec{x})$, or the logical constant $\mathit{false}$, and
$\mathit{body}$ is $q_1(\vec{x}_1) \land \ldots \land q_n(\vec{x}_n) \land \varphi(\vec{x})$ ($n \geq 0$), 
where $q_i$ are predicate symbols taking different subsets of $\vec{x}$ as arguments
and all functions and predicates in $\varphi$ are interpreted in a background theory.
For example, $\varphi$ could be of the form $x + 2 \cdot y > 4$, where 
the meaning of $+,\cdot, >$ are defined by the standard model for arithmetic.

The engines summarized in Table~\ref{tab:apps} apply to different classes of Horn formulas.
They also use widely different engines. These engines range from using interpolations to encode 
classes of failed SLD resolution proofs, bottom-up Datalog for finite domains using explicit hash-tables or using tables that are encoded symbolically,
to SLD resolution with tabulation. For finite domains, the Z3 Datalog engine supports
stratified negation. As a default table representation
it uses hash-tables with column indexes, which is suitable for 
domains where tables remain relatively small (up to a few million entries).
For networking, it encodes tables symbolically.
The idea with tabulation search is to establish that there are no derivations of
$\mathit{false} \leftarrow \mathit{body}$, but creating goals from the predicates in $\mathit{body}$, and carrying along the
side constraints from the interpreted formulas. Sub-goals that are found to be subsumed by previous sub-goals are discarded.

Methods that are suitable for software model checking are described in depth in~\cite{DBLP:conf/birthday/BjornerGMR15}.
Program analysis by reduction to logic program analysis has received steady attention from
the program analysis~\cite{DBLP:conf/cav/RepsSW04} and program verification 
communities~\cite{DBLP:conf/pldi/GrebenshchikovLPR12}. 
A common trait of these methods is that they look for \emph{symbolic} solutions to Horn clauses. In a nutshell,
a symbolic solution is a definition of the free predicates using interpreted formulas. For example, if
a CHC is of the form $\forall \vec{x},\vec{y} \ . \ p(\vec{x}) \leftarrow q(\vec{y}) \land \varphi(\vec{x},\vec{y})$, 
then a symbolic solution are formulas $\psi_p(\vec{x})$ and $\psi_q(\vec{y})$, such that the formula
$\forall \vec{x},\vec{y} \ . \ \psi_p(\vec{x}) \leftarrow \psi_q(\vec{y}) \land \varphi(\vec{x},\vec{y})$ is valid
modulo a background theory. Symbolic solutions correspond to inductive invariants from Hoare Logic.
As a general takeaway, we show that all standard transformations on Horn clauses, such as
Magic transformations, in-lining for K-indutiveness, 
partial assertion in-lining, and fold-unfold transformations
preserve inductive invariants modulo interpolation. 
That is, each transformation also comes with a (cheap) method for translating symbolic
solutions of the original Horn clauses to solutions of the transformed ones.
Conversely, solutions to the transformed systems
can be translated to solutions to the original clauses when 
the background theory admits Craig interpolation. 
Some transformation methods were invented in the context of symbolic model checking.
For example, K-induction is a widely used method where the invariant is shown to hold in the first $K$ steps, 
and then it is shown inductive by assuming it holds in the previous $K$ steps. Property Directed Reachability, PDR,
was invented for finite state machine (hardware) model checking, we imported it to Horn clauses with theories.
This allows going beyond finite state machines and model programs with procedure calls.
A clear appeal of using Horn clauses is that they offer a well defined interchange format between tool layers 
that handle details specific to a programming language in one end, and symbolic solving engines in the other end.


\mysec{Network Verification and Logic Programming}

Networking and logic programming are no strangers~\cite{DBLP:conf/sigcomm/LooHSR05,DBLP:conf/nsdi/KoponenABCCFGGIJLLLPPPRSSSTWYZ14}.
In the context of Z3, we developed a special purpose engine for verifying packet based forwarding. 
Our engine is called NoD (Network optimized Datalog).
NoD is currently used as part of the Batfish 
tool~\cite{DBLP:conf/nsdi/FogelFPWGMM15} that checks 
reachability properties in wide area networks that 
are configured using BGP and OSPF. One of the inherent challenges with using
distributed routing protocols, such as BGP and OSPF,
is that the routes are computed without reference to 
network access control lists (ACLs). NoD checks 
that the ACLs are placed in a consistent way.

The table representation in NoD represents sets of bit-vectors using don't care bits.
For example, the two bit-vectors $101$ and $111$ can be represented using just one
\emph{ternary} bit-vector $1{\star}1$. The representation also includes set-difference
operations, such that $10,00,11$ is represented by $11\setminus \{\star\star\}$.
The use of ternary bit-vectors for packet-switched network analysis was chosen because
the main operations performed on packets are bit-masking and updates to ranges of bits.
These operations easily map to set operations on ternary bit-vectors.

\mysec{Technological Barriers and Aspirations}
There are many possible extensions and improvements of Z3's Horn clause engines.
For symbolic model checking we found that a notion of \emph{model-based projection} 
to be useful. It amounts to partial quantifier elimination that use ground models as a starting point.
When coupled with proof-based half-interpolation it offers a powerful combination that applies
to several theories, such as integer, real arithmetic, arrays, polynomial real arithmetic, 
and algebraic data-types. There are many unexplored extensions to using model-based projection with
proof-based interpolation. Our engines for finite domains currently do currently not mix well with
engines that are suitable for infinite domains. Abstract interpretation offers hints of one approach, 
using \emph{reduced products}, but our current use of reduced products have not been successful in
meshing SMT with abstract interpretation approaches.



\pagebreak

\def\bibdir{tarau}




\mytitle{
  Logic Programming: from NLP to NLU?
}


\myauthor{
Paul Tarau, 
       {University of North Texas} 
}


Natural Language Processing (NLP) was one of the original motivations leading to programming in logic \cite{KOW79} back in the seventies, with  Colmerauer's Metamorphosis Grammars \cite{COL78} and with  Pereira and D.H.D Warren's Definite Clause Grammars \cite{PW80}, 
enhanced later with mechanisms for hypothetical reasoning \cite{DT97:AGNL}. Montague Grammars (implemented in Prolog by D.S. Warren \cite{dswarren83}) and  work by Veronica Dahl on defining logic representations for more realistic fragments of natural languages (including long distance dependencies and anaphora resolution) \cite{DAHL94} have all shown a penchant of Logic Programming towards the higher objectives of Natural Language Understanding (NLU).

After being long delayed (and partly frozen by the AI winter) recent progress in NLU promises to bring disruptive paradigm shifts 
in human-computer interaction and several directly and indirectly related industries.
In fact, hopes for more logic-based NLU are high again, partly due to the possibility of sharing successful technologies and  tools
with successful fields like deep-learning neural-networks, machine learning, statistical parsers
and graph-based NLP.

This brings us to the obvious question: what new role can logic programming play in this new context?

When trying to sketch an answer, after more than a decade spent on other research topics, we became aware that one can  benefit today from the extended logic programming ecosystem, consisting of classic tools like Prolog, constraint solvers, Answer-Set Programming and SAT/SMT systems, as well as machine learning techniques implemented on top of inductive and probabilistic logic programming \cite{probLog2}.

One of the research directions we have worked on in the past, that turned out to be remarkably successful, is graph-based NLP. It has originated in a Prolog program that was using WordNet's semantic links to improve word-sense disambiguation (WSD) \cite{coling04:pr}, by building a graph connecting words, sentences with their semantic equivalence classes (synsets) and then guessing the most likely sense associated to a word, by running  the PageRank algorithm \cite{page98pagerank} on the graph. A few months later, an unsupervised version of the algorithm \cite{EMNLP:TR} based on graphs connecting sentences to word occurrences has been shown to extract high quality summaries and keywords. It later became one of the most popular techniques for the task \cite{radabook}, implemented in virtually all widely used programming languages and several NLP libraries.

While recently revisiting the topic, it became clear that involving logic programming tools is likely to enhance graph based NLP. The use of logic-based implementations of Combinatorial Categorial Grammars (CCGs) looks especially appealing as it provides lexicalized representations easier to correlate with words and word phrases. The NLU-component coming from extracting logic representations is likely to make the links in the graph structure more meaningful. 

A simple reducer for a subset of CCG rules looks as follows:

{\small
\begin{verbatim}
:-op(400,xfx,(/)).
:-op(400,xfx,(\)).

red(Xs):-red(Xs,s). % reduce a sentence to root symbol s.

red([S],S).
red([X/Y,Y|Xs],S):-red([X|Xs],S).
red([Y,X\Y|Ys],S):-red([X|Ys],S).
red([X/Y,Y/Z|Xs],S):-red([X/Z|Xs],S).
red([Y\Z,X\Y|Xs],S):-red([X\Z|Xs],S).
\end{verbatim}
}
Interestingly, Prolog's DCGs can be used to build the CCGs representation of a sentence as in: 
{\small
\begin{verbatim}
the -->[np/n].  
cat -->[n].
chased -->[(s/np)\np].
dog -->[n].
playful --> [n/n].
quiet -->[n/n].
quick -->[n/n].
and --> [X/X].

sent-->the,quick,and,playful,dog,chased,the,quiet,cat.
\end{verbatim}
}
When executed it accepts a sentence as follows:
{\small
\begin{verbatim}
?- sent(S,[]),red(S).
S = [np/n, n/n, n/n, n/n, n,  (s/np)\np, np/n, n/n, n] .
\end{verbatim}
}
More elaborate parsers can be built using CYK or $A^{*}$ parsers and tools like tabling in Prolog provide the means to do that efficiently.
Interestingly, the CCG parsing problem can  also be nicely expressed and executed directly with Answer Set Programming tools
as shown in \cite{aspccg12}.

Tools like the {\tt boxer} program \cite{boxer15} can, in combination with a statistically trained CCG parser like \cite{cc07},
build Prolog clauses describing semantically labeled first-order formulas representing input sentences, ready to be further
explored with standard and probabilistic logic programming algorithms as well as graph based algorithms exploiting the richer
link structure between their underlying concepts.

An emerging field in NLP these days is sentiment analysis, as knowing what the opinion of an author is about a topic is as important and knowing what a document is about. Modalities and negation detection provided by a logic component combining syntactic and semantic parsing can improve sentiment analysis. Figuring out the implicit entailment links important for understanding a story line or the rhetorical structures involved in an argument is also likely to benefit from logic representations.

Another NLU-minded application we have worked more than a decade or ago is the use of Prolog-based natural language-enabled agents \cite{lm}. 
They interacted with the Prolog version of WordNet and Google's metasearch API to bring in knowledge distributed over the internet.
The integration of logic inferences and a Prolog representation (as dynamic clauses or backtrackable assumptions \cite{DT97:AGNL}) of the agents' short-term memory, have 
significantly enhanced the quality of the dialog, with shared virtual worlds and interactive story telling systems developed
on top of them \cite{tidse:vista,tidse:wnet}. These days, fields like interactive story telling have become an integral part of computer games (e.g., Minecraft Story Mode) and voice-enabled software agents are part of major mobile phone (e.g., Siri, Cortana, Google Ok) and platforms are making their way in home automation systems (e.g., Alexa) and more generally in the upcoming Internet-of-Things (IOT) platforms. 

Revisiting some of the logic programming-based NLP tools we have used in the past, can today benefit from access to improved metasearch as well as massive online knowledge repositories like Wikipedia. Involving constraint programming libraries, now part of most widely used Prolog systems is  likely to improve the speed and the accuracy of WSD, an important NLU component. Involving SAT-solvers and ASP-based systems can help narrowing down some of the heavily combinatorial aspects related to the inherent ambiguity of natural language, as well as in dealing with incomplete or noisy information streams one faces in voice and image recognition tasks.

Finally this brings us to the possible synergies between logic programming and deep-learning neural network technologies \cite{deep15}, credited for the new ``AI-Spring'', brought by successful applications to popular fields like vehicle automation, vision and internet search. Tools like Google's TensorFlow and word2vec \cite{mikolov14} are specifically focused on enabling extensions transforming quantitatively represented meaning fragments into more human-friendly logic representations, ready for  inference steps that reveal implicit connections between facts and events.
Integrating logic programming components with this family of tools, possibly involving the quantitative means provided by probabilistic logic programming opens the door for being part of this the re-emergence of AI-based techniques in new application domains. 

Of special practical interest are logical formalisms based on lexicalized natural language representations (e.g., CCGs) that are likely to enable interaction at word level between symbolic and connectionist representations. 
As the same lexicalized representations can also enable  synergies with  graph-based methods in natural language processing \cite{radabook}, we expect a significant practical impact from  logic programming tools bringing together these NLP fields.



\end{document}